\documentclass[aps, prb, twocolumn, showpacs, preprintnumbers, amsmath, amssymb,
  floats, floatfix, longbibliography, 10pt]{revtex4-1}
\usepackage{graphicx, xcolor, hyperref, multirow, nicefrac, mathtools}
\usepackage{soul}
\setulcolor{red}
\setstcolor{red}
\sethlcolor{yellow}

\newcommand{\ph}{\hphantom{0}}

\begin{document}

\title{Entanglement transition from variable-strength weak measurements}

\author{M. Szyniszewski}
\email{mszynisz@gmail.com}
\affiliation{Department of Physics, Lancaster University, Lancaster LA1 4YB,
  United Kingdom}

\author{A. Romito}
\affiliation{Department of Physics, Lancaster University, Lancaster LA1 4YB,
  United Kingdom}

\author{H. Schomerus}
\affiliation{Department of Physics, Lancaster University, Lancaster LA1 4YB,
  United Kingdom}

\date{\today}

\begin{abstract}
We show that weak measurements can induce a quantum phase transition of
interacting many-body systems from an ergodic thermal phase with a large entropy
to a nonergodic localized phase with a small entropy, but only if the
measurement strength exceeds a critical value. We demonstrate this effect for a
one-dimensional quantum circuit evolving under random unitary transformations
and generic positive operator-valued measurements of variable strength. As
opposed to projective measurements describing a restricted class of open
systems, the measuring device is modeled as a continuous Gaussian probe,
capturing a large class of environments. By employing data collapse and studying
the enhanced fluctuations at the transition, we obtain a consistent phase
boundary in the space of the measurement strength and the measurement
probability, clearly demonstrating a critical value of the measurement strength
below which the system is always ergodic, irrespective of the measurement
probability. These findings provide guidance for quantum engineering of
many-body systems by controlling their environment.
\end{abstract}

{\maketitle}

\section{Introduction}

Generic many-body quantum systems obey the eigenstate thermalization hypothesis
(ETH), according to which they establish the characteristics of thermal
equilibrium in a finite time, leading to ergodic dynamics in local
observables~\cite{Deutsch1991, Srednicki1994, DAlessio2016, Borgonovi2016}.
These systems are characterized by highly entangled eigenstates, i.e., states
following an extensive scaling of their entanglement entropy, a behavior known
as the volume law. However, systems with local interactions can display a
transition to a phase in which the entanglement entropy obeys a subextensive
area-law scaling, where ETH is violated. A paradigmatic case is that of
many-body localization~\cite{Basko2006, Gornyi2005, Altman2015,
Nandkishore2015}, in which the entanglement transition~\cite{Znidaric2008,
Bauer2013, Kjall2014, Luitz2015} is driven by the strength of a local disordered
potential. The fate of this transition in open quantum systems is the subject of
intense current investigations~\cite{Louschen2017, Xu2018, Rieder2018,
Marino2018, Vakulchyk2018}. A promising path to address the entanglement
transition in this setting is to consider a randomly driven system on which the
external environment acts as a quantum detector~\cite{Li2018, Chan2018,
Skinner2018, Li2019}. In this scenario, the free unitary evolution of the
unobserved system leads to a ballistic temporal growth of
entanglement~\cite{Nahum2017, Keyserlingk2018, Nahum2018}, until it settles into
a highly entangled quasi-stationary state that follows the volume law. This
tendency is counteracted by local measurements, which induce a stochastic
nonunitary backaction. Very recently it has been shown that when local
projective measurements are performed frequently enough one encounters an
entanglement transition to a quasi-stationary state characterized by an area
law~\cite{Li2018, Skinner2018, Chan2018}.

Projective measurements capture an important but restricted class of
environments. A much larger class can be accessed by adopting a more general
perspective, according to which quantum measurements are not necessarily
characterized by discrete projections in the time evolution, but are generically
given by positive operator-valued measurements (POVMs) with a continuous
spectrum of outcomes and a corresponding stochastic back-action onto the
system~\cite{Jacobs2014, Wiseman2009, Clerk2010}. The measurement strength
$\lambda$ can then be controlled to continuously interpolate between the
unobserved case and projective measurements. Such weak measurements are also
routinely employed in experiments for quantum states readout~\cite{Clerk2010}
and quantum feedback protocols~\cite{Vijay2012, Blok2014, Lange2014}, and can be
used as a theoretical tool to reproduce Lindblad dynamics of open systems via
trajectory averaging~\cite{Jacobs2014, Wiseman2009}. They, therefore, describe a
much more versatile framework to model the influence of the environment on an
open quantum system. The effect of POVM on entanglement evolution has been 
addressed for continuous measurements in free fermionic systems \cite{Cao2018}, 
while the existence of an entanglement transition driven by measurement 
strength for interacting systems has been inferred by the study of mutual 
information~\cite{Li2019}. 

In this paper, we systematically study the effect of these weak measurements on
the ergodic properties of randomly driven many-body systems, modeled as a
quantum circuit with local interactions. By analyzing the entanglement entropy
and its variance we show that weak measurements can indeed drive a transition to
a nonergodic phase obeying the area law. Similarly to the case of projective
measurements~\cite{Li2018}, the result resembles that of the Zeno effect in
which a transition to frozen dynamics is obtained by sufficiently frequent
projective measurements. Importantly, we identify a minimal critical measurement
strength below which no localization is possible regardless of the measurement
probability $p$. These results are obtained by mapping out the phase boundary by
two different methods, based on the data collapse and scaling of the entropy for
different system sizes, as well as the statistical fluctuations of the entropy,
which turn out to be maximal at the phase transition. Beyond providing this
specific phenomenology, our results open up a broader avenue to study many-body
quantum dynamics in open systems, understand their properties in more detail,
and guide the implementation of quantum feedback control for technological
applications.

\section{Model}
\begin{figure}[t]
	\centering
	\includegraphics[width=\linewidth]{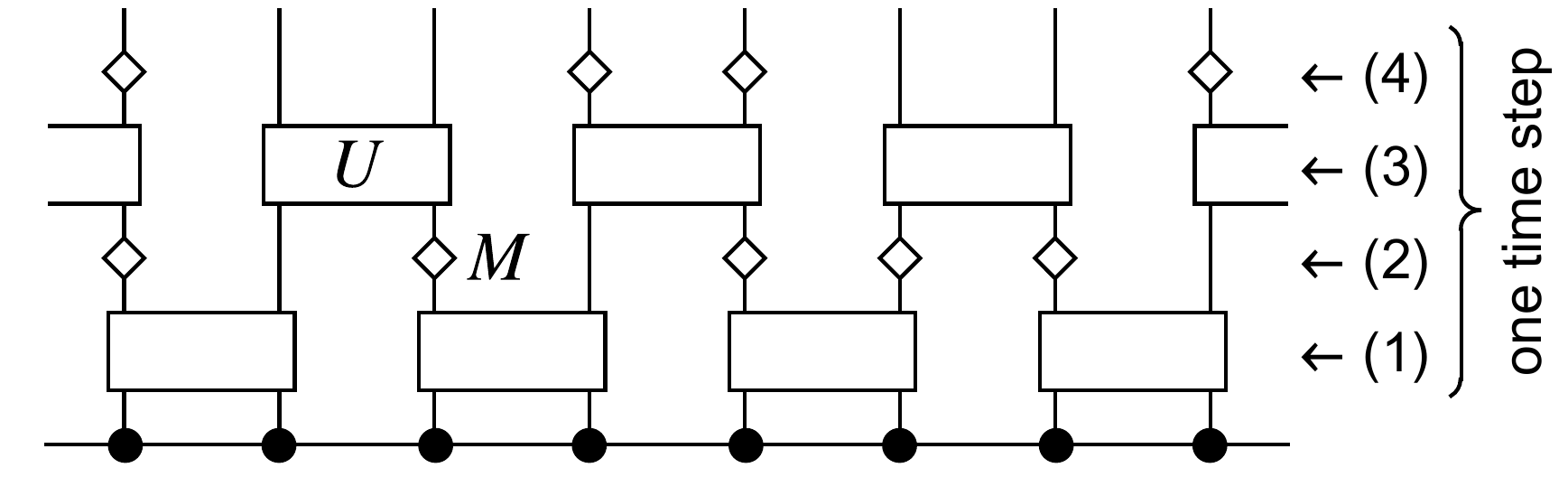}
	
	\caption{
  Diagrammatic representation of the evolution of a weakly measured quantum
  circuit during one time step. The dots represent the locations of $L$ sites
  with spin 1/2. The rectangular blocks correspond to unitary operators $U$ on
  two neighboring spins, while the diamonds correspond to nonunitary evolution
  induced by the weak measurements $M$ that occur with probability $p$ and have
  strength $\lambda$. The displayed configuration of diamonds denotes one
  possible realization of the measurement locations. We consider the dynamics of
  the system commencing from an unentangled N\'eel state.}
		
  \label{fig:evolution}
\end{figure}

We analyze the entanglement transition in a model consisting of a spatially
periodic one-dimensional quantum circuit consisting of a chain of $L$ spins,
where  $L$ is assumed to be even to allow the partition of the chain into two
subsystems of equal length. The chain evolves in time under a sequence of
discrete time steps that are inherently stochastic and consist of a sequence of
four operations, as schematically depicted in Fig.~\ref{fig:evolution}. First,
unitary two-spin operators $U$ are applied between each odd site and the next
neighboring site. Each operator is chosen randomly and independently according
to the Haar measure over the set of unitary operations for two spins. Second, a
measurement $M$ is carried out on each site with probability $p$. In this work,
$M$ is a properly normalized Kraus operator (not necessarily a projector)
associated with a POVM~\cite{Nielsen2010, Peres2002, Brun2002, Jordan2006,
Jacobs2014},  as specified in detail below. Third, analogous to the first step,
random unitary two-spin operators $U$ are applied between each even site and the
subsequent odd site. Finally, another set of single-site measurements $M$ is
carried out, with the same probability $p$ as in the second step. These unitary
operators and measurements vary throughout space and from time-step to
time-step, rendering the local interactions and measurements disordered and
aperiodic. Therefore, both the unitary evolution and the measurement operations
are of a stochastic nature.

We implement specific POVM operations that model the detection of $z$-component
of spin via coupling to an external pointer. While the results are not expected
to depend strongly on the specifics of the POVM model, the studied example
provides a physically intuitive picture of the process. The basic idea is to
describe the detector as a pointer of coordinate $x$ and conjugate variable $p$,
$[x,p]=i$, that interacts with the system during a time interval $\delta t$ with
a Hamiltonian of the form $H_{\rm int}=\lambda \sigma_z p$, where $\sigma_z$ is
the observable being measured. Since the interaction Hamiltonian implies
$\dot{x}=\lambda \sigma_z$, the dynamics of the pointer position tracks the
observable of interest~\cite{vonNeuman1938, Caves1987, Das2014}. For a pointer
initially prepared in a Gaussian state of width $\Delta$, $G_\Delta(x) =
\exp(-x^2/2\Delta^2)/(\pi^{1/4}\Delta^{1/2})$, the probability distribution of
outcome $x$ following the measurement is
\begin{equation}
	P(x) = G^2_\Delta (x-\lambda) \langle\psi| \Pi_{+}^{(j)} |\psi\rangle
	+ G^2_\Delta (x+\lambda) \langle\psi| \Pi_{-}^{(j)} |\psi\rangle,
	\label{eq:measurement_prb}
\end{equation}
where $\vert \psi \rangle$ is the state of the system before the measurement and $\Pi_{\pm}^{(j)} =
(1\pm\sigma_{z}^{(j)})\bigotimes_{i \neq j} \boldsymbol{1}^{(i)}$ projects the
$j$-th site component of the state onto the spin-up or spin-down subspace. Conditional to the detector reading $x$, the state of the system is updated as
\begin{equation}
  |\psi'\rangle = {1\over\sqrt{P(x)}} \left[
  G_\Delta (x-\lambda) \Pi_{+}^{(j)} |\psi\rangle
  + G_\Delta (x+\lambda) \Pi_{-}^{(j)} |\psi\rangle
  \right].
\label{eq:measurement_new_state}
\end{equation}
The effect of the measurement depends on the measurement strength $\lambda
/\Delta$ \cite{Caves1987, Aharonov1988}. For $\lambda \ll \Delta$,
Eq.~(\ref{eq:measurement_prb}) reproduces the binomial distribution of a strong
projective measurement of $\sigma_z^{(j)}$ with peaks at $x=\pm \lambda$ of
heights proportional to $\langle\psi| \Pi_{\pm}^{(j)} |\psi\rangle$. In the
opposite limit $\lambda \ll \Delta$, we have a  weak measurement with a broad
Gaussian distribution of $x$ that does not distinguish between spin up and spin
down in a single measurement shot and induces a back-action perturbatively small
in $\lambda / \Delta$.

The evolution of the random circuit from the measurements at generic value of
$\lambda/\Delta$ is a stochastic process determined by the outcome of the
detector reading $x \in\mathbb{R}$ drawn form the distribution in
Eq.~(\ref{eq:measurement_prb}) and consequent update of the state from
Eq.~(\ref{eq:measurement_new_state}). Note that, in the continuous measurement
limit where a sequence of such measurements is taken at intervals $\delta t \to
0$ with $\lambda=\lambda_0 \sqrt{\delta t}$, the model generically reproduces a
continuous Lindblad equation~\cite{Jacobs2006}, and for our case reduces to the
measurement model used in Ref.~\onlinecite{Cao2018}.

\section{Volume to Area law transition}

To discriminate between ergodic and nonergodic many-body phases we utilize the
bipartite entanglement entropy
\begin{equation}
  S = -\text{tr} (\rho_A \ln \rho_A),
  \label{eq:entropy-definition}
\end{equation}
where $\rho_A$ is the reduced density matrix of a subchain $A$. This entropy is
extensively employed to quantify the entanglement in quantum systems, and its
scaling properties with length can be used to determine whether the system is in
an ergodic thermal-like phase or a nonergodic localized
phase~\cite{Znidaric2008, Eisert2010, Bardarson2012, Bauer2013, Kjall2014,
Luitz2015, Chan2018}. Alternative entanglement measures used in the literature
include R\'enyi entropies~\cite{Li2018, Skinner2018, Li2019} and mutual
information~\cite{Eisert2010, Skinner2018, Li2019}. We focus on the case where
the subchain $A$ is of length $L/2$, which corresponds to cutting the circuit
into two halves of equal size. Note that a cut may be commensurate or
incommensurate with the unitary operations $U$ in each layer of the circuit,
which can result in differences between systems of length $L=4k$ or $L=4k+2$.
We, therefore, place these cuts randomly.

\begin{figure}[t]
	\centering
	\includegraphics[width=\linewidth]{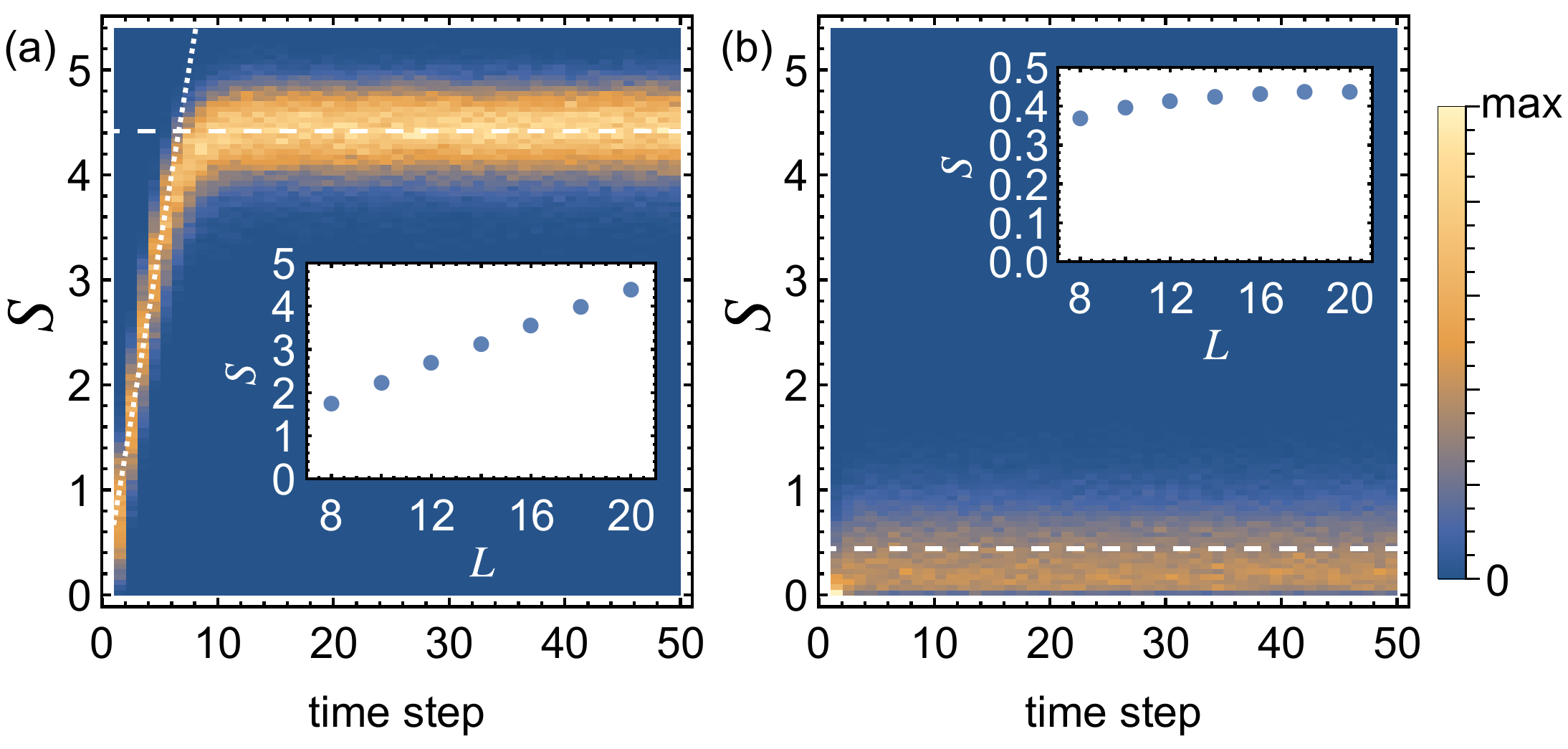}
	\caption{
  Color-coded histogram of the bipartite entanglement entropy $S$ for a quantum
  circuit of length $L=20$ for representative parameters in two regimes, with
  measurement strength (a) $\lambda / \Delta = 0.2$ and (b) $\lambda / \Delta =
  0.6$ while in both cases the measurement probability is $p = 0.9$. The dashed
  lines indicate the averaged quasistationary values of the entropy after the
  initial transient growth. The insets show the dependence of these
  quasistationary values with system size. According to these results, in (a)
  the system is ergodic and obeys a volume law, while in (b) it is nonergodic
  and follows an area law. Each panel contains data from 1000 realizations.}

	\label{fig:entro-evol}
\end{figure}

The stochastic evolution of the entropy over time is illustrated in
Fig.~\ref{fig:entro-evol}. The figure displays a color-coded histogram obtained
from 1000 realizations of the stochastic evolution commencing from an
unentangled N\'eel state, and contrasts two cases of different measurement
strength while the measurement probability is fixed to $p=0.9$. For the weaker
measurement strength $\lambda/\Delta=0.2$ [panel~(a)] the entropy increases
ballistically over $\sim L/2$ time steps and then saturates at a large value,
which increases linearly with the system size (see the inset). This is the
signature of an extensive entropy scaling according to a volume law. The entropy
in the quasistationary regime at long times is close to the value in an
unobserved system ($\lambda=0$ or $p=0$), where the average entropy is predicted
to be given by~\cite{Page1993}
\begin{equation}
  \langle S\rangle = {L\over 2} \ln 2 - {1\over 2}.
  \label{eq:average-entropy}
\end{equation}
A qualitatively different behavior is observed for the stronger measurements
with $\lambda / \Delta =0.6$  [panel~(b)]. There the entropy saturates quickly,
and at a value independent of the system size (see the inset). This corresponds
to an area law, as encountered in a localized phase.

\begin{figure}[t]
	\centering
	
	\includegraphics[width=\linewidth]{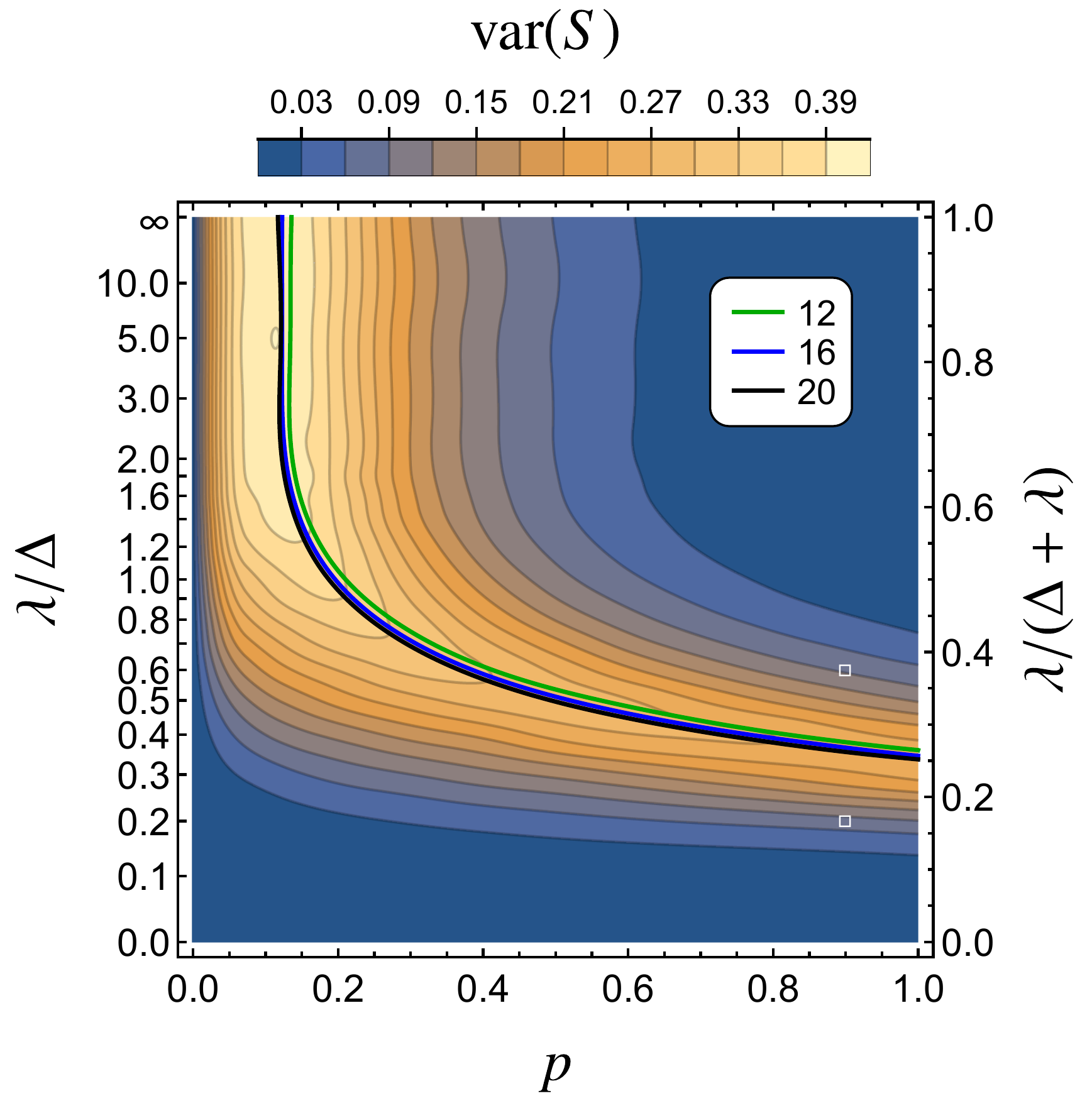}
	
	\caption{
  Color-coded variance $\mathrm{var}(S)$ of the entanglement entropy for varying
  measurement probability $p$ and the measurement strength $\lambda/\Delta$ in a
  system of length $L = 20$. We observe two regions of small entropy
  fluctuations, separated by a ridge where the fluctuations are large. The
  superimposed thick lines show the location of the maximal variance for system
  sizes $L = 12, 16, 20$, which serves as an indication for the phase boundary
  (see also the finite-size scaling in Fig.~\ref{fig:p1_entropy_var}). The two
  square markers show the two representative points selected for
  Fig.~\ref{fig:entro-evol}.}
	
	\label{fig:phase_space}
\end{figure}

The observation of two phases suggests that a phase boundary exists in the
parameter space spanned by $p$ and $\lambda / \Delta$. A clear indication of the
support of the phases can be obtained from the variance of the entropy, shown in
Fig.~\ref{fig:phase_space}. We again account for 1000 realizations and collect
the data from the quasistationary regime (from $\sim L/2$ to a cutoff of $100$
time steps, much larger than the range of the transient entropy growth for all
systems sizes studied here). As before, the initial state is an unentangled
N{\'e}el state. We observe that there are two regions where the entropy
fluctuations are small, separated by a transition region with a pronounced
increase of the fluctuations. The contour of maximal variance (thick lines) is
stable for different system sizes, while the critical fluctuations along this
contour increase with system size, as we further analyze below. The contour
indicates that there exists a critical value  $p_\text{crit}$ for the
measurement probability below which there are too few measurements to localize
the system, as previously identified in analogous models based on projective
measurements (hence $\lambda/\Delta\to\infty$)~\cite{Li2018, Chan2018,
Skinner2018}. Departing from the projective measurement scenario, the
measurement probability at the transition becomes dependent on the measurement
strength, so that measurements have to occur more frequently when the
measurement strength decreases. However, even for permanent measurements
($p=1$), the transition occurs at a finite measurement strength, which we denote
as $(\lambda/\Delta)_\text{crit}$. Below this measurement strength, the system
is always in the ergodic phase. This is our main result. In the remainder of
this paper we characterize these transitions in detail, and in particular,
confirm that they become sharp in the limit of a large system size.

\begin{figure}[t]
	\centering
	\includegraphics[width=\linewidth]{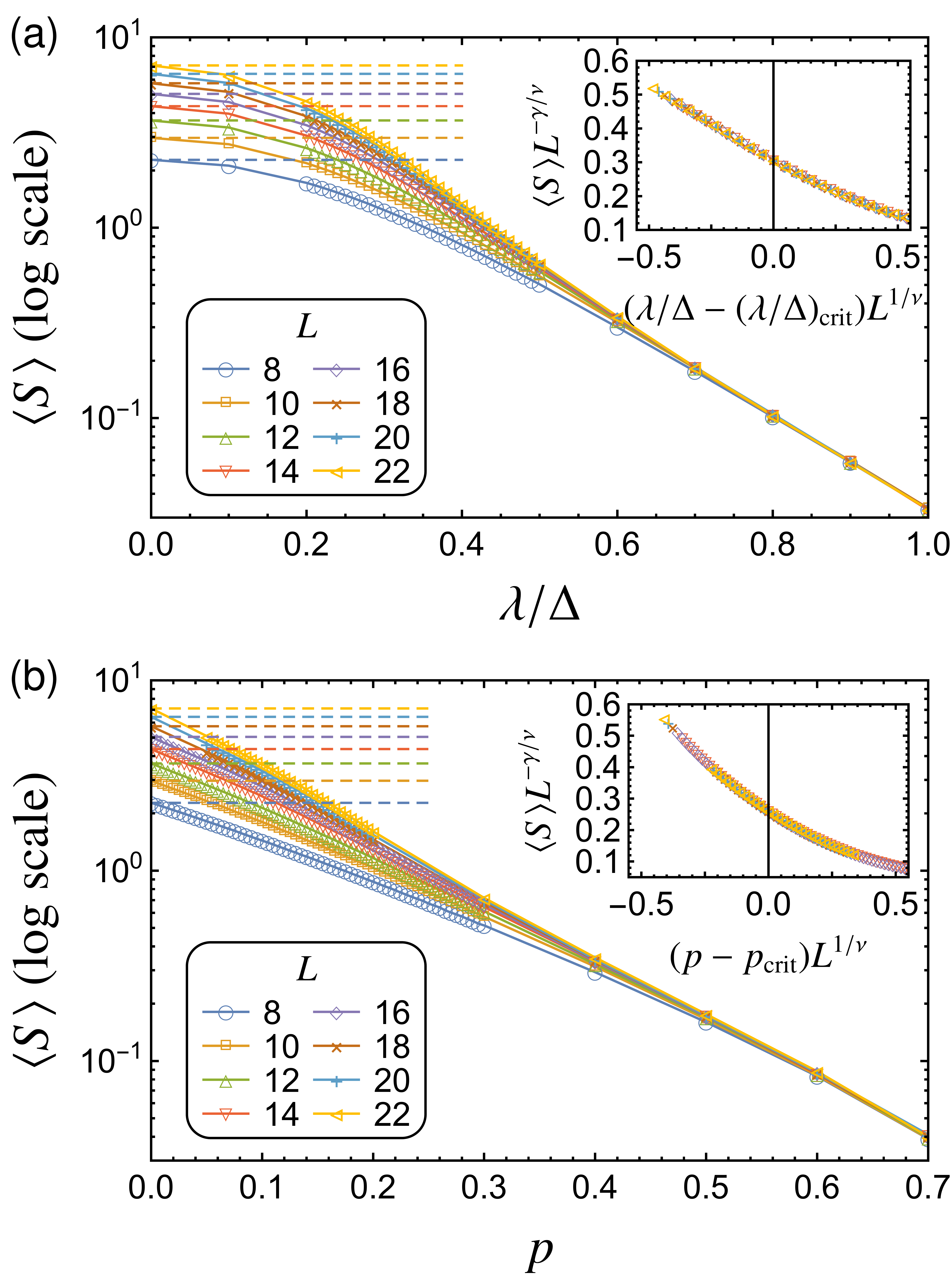}
	
	\caption{
  Comparison of the averaged entropy $\langle S\rangle$ for different system
  sizes $L$ across the phase transitions with (a) $p=1$ (where measurement are
  always performed), and (b) $\lambda/\Delta=10$ (where measurements are always
  strong). The dashed lines show the analytical prediction
  (\ref{eq:average-entropy}) for an unobserved system. The insets show the
  finite-size scaling for circuits of length $L \ge 14$. In (a), the data
  collapse gives the critical measurement strength $(\lambda/\Delta)_\text{crit}
  = 0.30(1)$ with critical exponents $\gamma = 1.38(7)$, $\nu = 1.96(2)$, while
  in (b) we obtain the critical measurement probability $p_\text{crit} =
  0.110(3)$ with $\gamma = 1.94(2)$, $\nu = 2.352(5)$. Statistical errors are
  smaller than the marker size. }
	
	\label{fig:p1_entropy}
\end{figure}

\section{Critical behavior}

In order to determine the two critical parameters, we consider two distinct
transition scenarios: (a) the case where the measurement is always performed
($p=1$) while the measurement strength $\lambda/\Delta$ is varied, and (b) the
case of almost-projective measurements with $\lambda/\Delta=10$ and varying
probability $p$. Fig.~\ref{fig:p1_entropy} displays the average $\langle
S\rangle$ of the entanglement entropy for both scenarios. For small $\lambda /
\Delta$ or $p$ the entropy clearly follows a volume law, as can be seen by
comparison with the theoretical prediction \eqref{eq:average-entropy} for an
unobserved ergodic system. At large values of these parameters, however, the
entropy becomes independent of the system size, so that an area law is observed.
That we indeed deal with a well-defined transition follows from the finite-size
scaling analysis~\cite{Fisher1971, Fisher1972} shown in the insets. For this, we assume that in the critical region the
correlation length scales as $\xi \sim |x - x_\text{crit}|^\nu$ and the entropy
scales as $\langle S\rangle \sim |x - x_\text{crit}|^\gamma$, where $x=p,
(\lambda/\Delta)$ is the varied parameter. The scaling ansatz then takes the
form
\begin{equation}
  \langle S\rangle L^{-\gamma/\nu} = F\left(L^{1/\nu} (x-x_\text{crit})\right),
  \label{eq:fss}
\end{equation}
where $F$ is an unknown scaling function. The transition point $x_\text{crit}$
and the critical exponents $\nu$ and $\gamma$ can be obtained by means of a data
collapse~\cite{Serra2000, Beach2005}. The data collapse shown in the insets then
gives the values $(\lambda/\Delta)_\text{crit} = 0.30(1)$ and $p_\text{crit} =
0.110(3)$.

\begin{figure}[t]
  \centering
  \includegraphics[width=\linewidth]{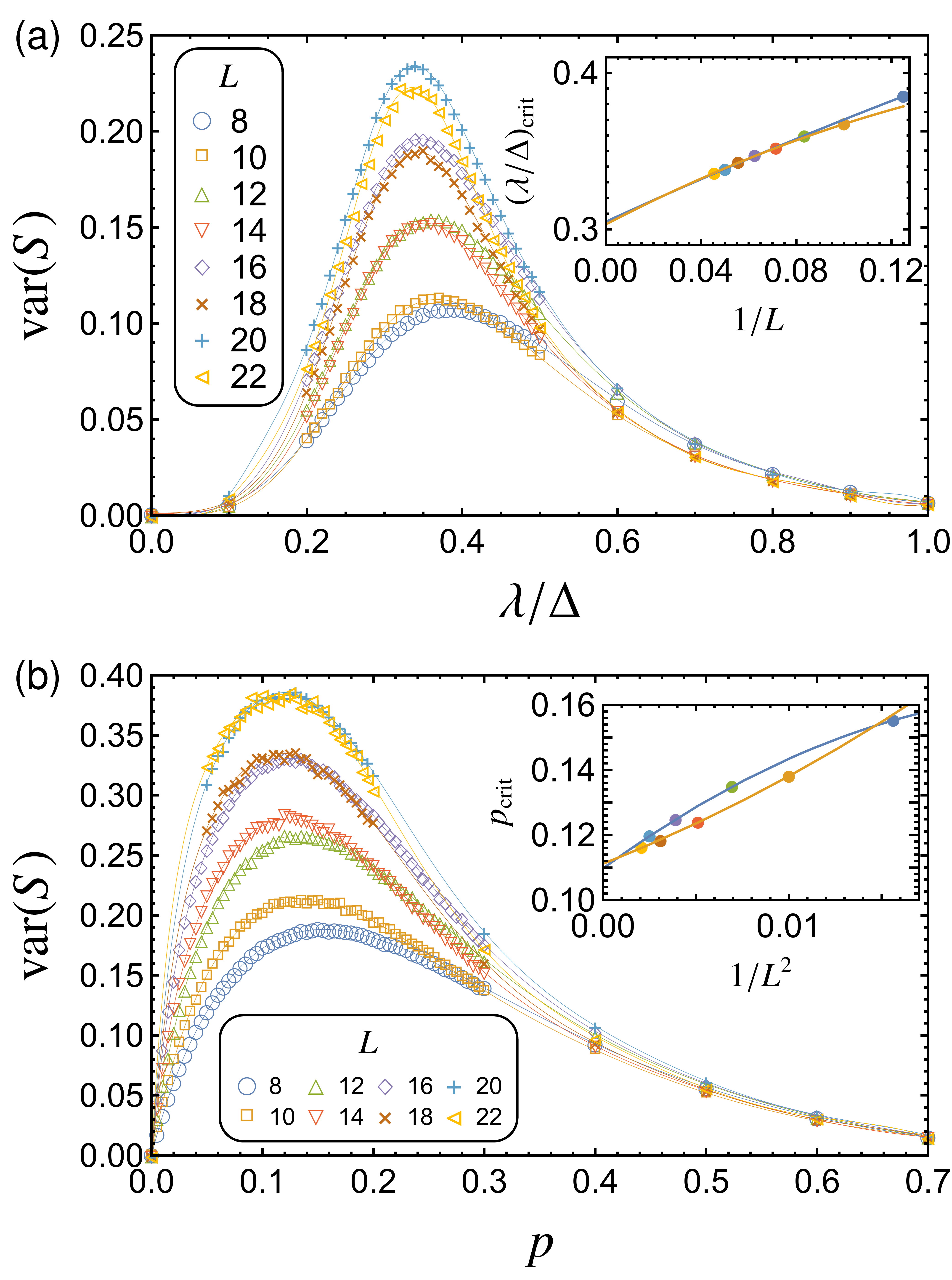}

  \caption{
  Comparison of the variance $\mathrm{var}(S)$ for different system sizes $L$
  across the same phase transitions as in Fig.~\ref{fig:p1_entropy}. The thin
  lines are fits that allow reliable extraction of the maxima. The insets show
  the finite-size scaling of the position of the maxima, where the dark blue
  line represents the extrapolation for chains of length $L=4k$, while the light
  orange line represents the extrapolation for lengths $L=4k+2$. Statistical
  errors are smaller than the marker size.}

  \label{fig:p1_entropy_var}
\end{figure}

Fig.~\ref{fig:p1_entropy_var} displays analogous results for the entropy
fluctuations captured by the variance $\mathrm{var}(S)$. The main panels present
the numerical data, along with a fit utilized to enable reliable extraction of
the position and height of the maximal variance. We observe that the positions
of the maxima display a remarkably small drift with the system size, which
reliefs us from the complications encountered in other cases~\cite{Kjall2014,
Bera2015}. Therefore, we can extract the phase transition directly from
finite-size scaling of the locations of the maxima, which is again presented in
the insets. We observe two scaling curves, one originating from systems of
length $L=4k$ and another one from systems of length $L=4k+2$, which, as
mentioned before, differ by the commensurability of the bipartition in each
layer. Reassuringly, both curves extrapolate to consistent transition points for
$L\to\infty$, giving the values listed in Table~\ref{tab:critical_params}.
Combining both transition curves into a simultaneous extrapolation we then
obtain the values $(\lambda / \Delta)_\text{crit} = 0.304(3)$ and $p_\text{crit}
= 0.1103(7)$, in excellent agreement with the values found from the average
entropy data collapse.

\begin{table}[t]
  \begin{tabular}{lcc}
    \hline\hline
                            & \multicolumn{2}{c}{Critical values}\\
    \cline{2-3}
    Method & $(\lambda / \Delta)_{\text{crit}}$ & $p_{\text{crit}}$\\
                            & $(p = 1)$    & $(\lambda / \Delta = 10)$\\
    \hline
    Data collapse           & $0.30(1)$\ph & $0.110(3)$\ph\\
    Extrap. ($L = 4 k$)     & $0.305(5)$   & $0.1097(6)$\\
    Extrap. ($L = 4 k + 2$) & $0.303(4)$   & $0.1112(7)$\\
    Simultaneous extrap.    & $0.304(3)$   & $0.1103(7)$\\
    \hline\hline
  \end{tabular}

  \caption{Critical values of the measurement strength $\lambda/\Delta$ and the
  measurement probability $p$ as obtained from the finite-size scaling in
  Figs.~\ref{fig:p1_entropy} and \ref{fig:p1_entropy_var}.}

  \label{tab:critical_params}
\end{table}

The choice of the scaling function and the corresponding critical exponents are
corroborated by the analysis of a different (logarithmic) scaling function and
higher R\'enyi entropies, as reported in Appendix~\ref{app:scaling}. The
logarithmic scaling does not seem to provide a better fit than the power law in
Eq.~(\ref{eq:fss}). A power law scaling ansatz for the higher order R\'enyi
entropies is consistent with the results in Table~\ref{tab:critical_params}.

\begin{figure}[t]
  \centering
  \includegraphics[width=\linewidth]{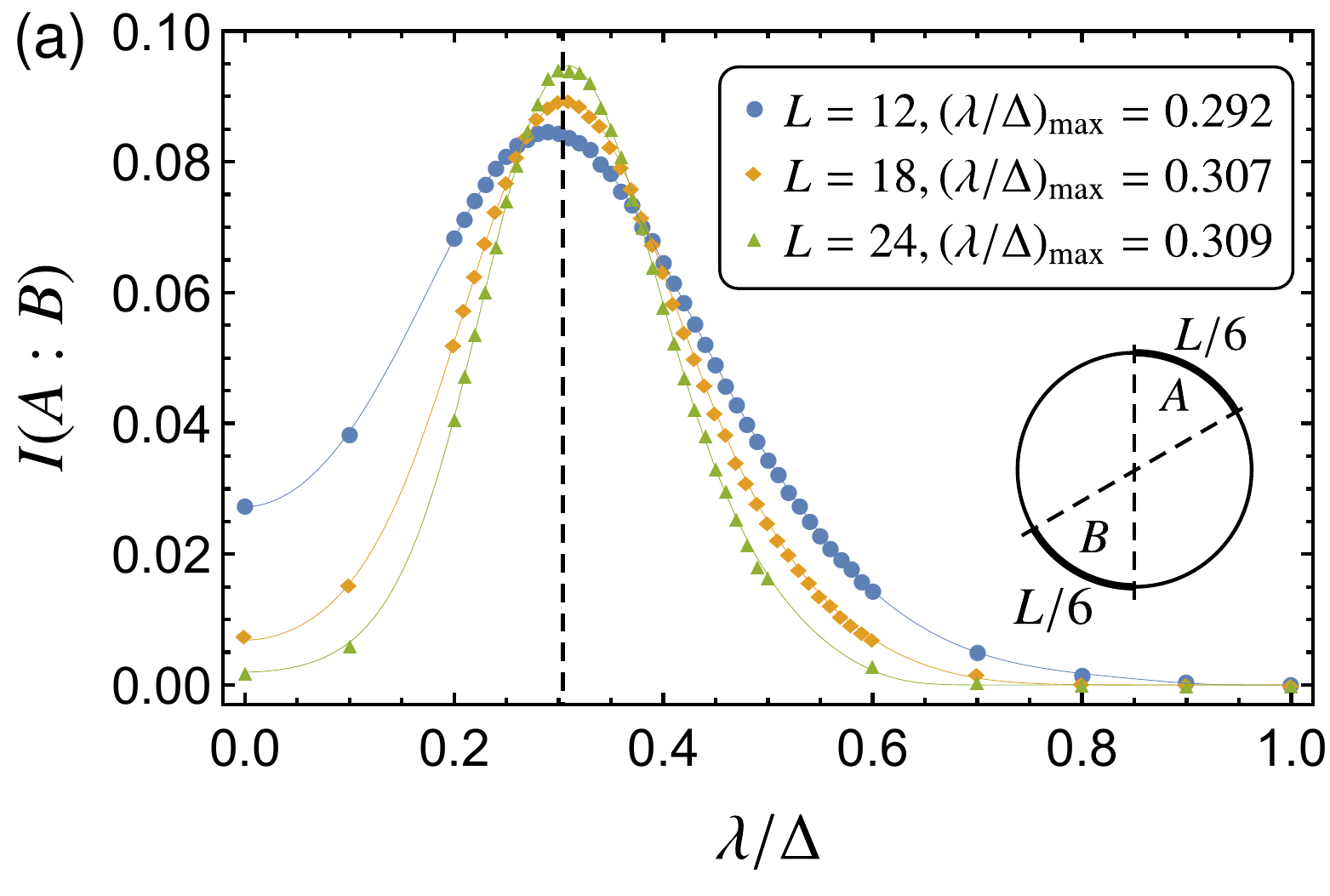}
  \includegraphics[width=\linewidth]{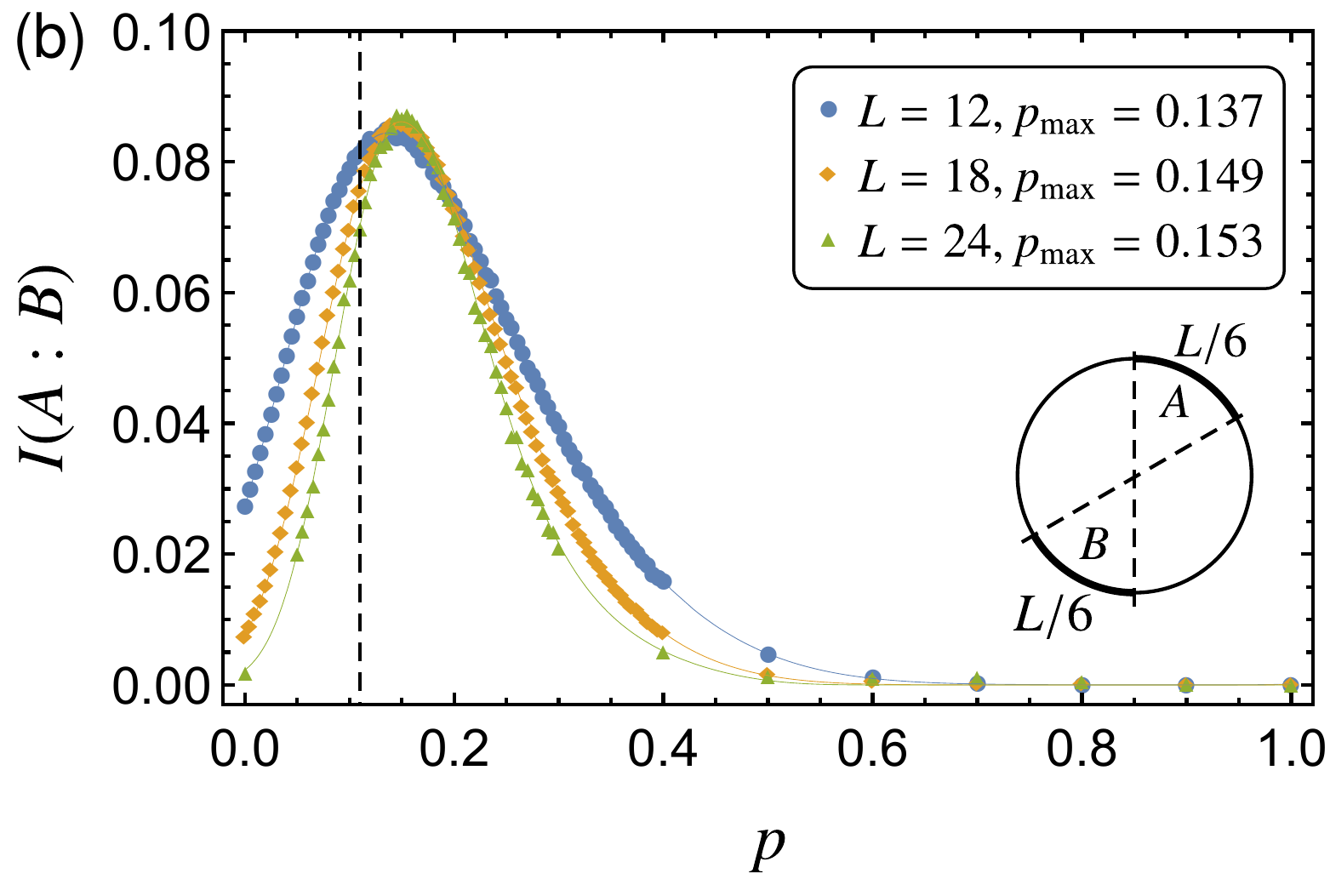}

  \caption{Behavior of the mutual information $I(A:B)$ as one increases the
  system size for the transitions with (a) $p = 1$ and (b) $\lambda/\Delta =
  10$. Regions $A$ and $B$ were chosen be antipodal and of length of $L/6$.
  Dashed lines indicate the transition values predicted in
  Table~\ref{tab:critical_params}.}

  \label{fig:mutual_information}
\end{figure}

Our results are also supported by the behavior of the mutual information
$I(A:B)$, which is defined as $I(A:\nobreak B) = S(A) + S(B) - S(A\cup B)$ for
two regions $A$ and $B$, and is a measure of correlations between subsystems in
the chain. We analyze $I(A:B)$ for antipodal regions $A$ and $B$ on a periodic
system of size $L$ for different $L$, keeping sizes $L/6$ of regions $A$ and $B$
constant relative to the system size.
Fig.~\ref{fig:mutual_information} shows that the mutual information exhibits a
clear peak near the transition predicted in
Table~\ref{tab:critical_params} (dashed lines).

\section{Conclusions}

In conclusion, weak local measurements can drive open many-body systems into a
nonergodic low-entropy phase, but only if the measurement strength exceeds a
critical value. This means that a continuously observed many-body  system can
remain ergodic even if the observation strength is finite. We demonstrated our
findings  for a quantum circuit, evolving under random local unitary operations
and local positive operator-value measurements modeled as a  Gaussian probe, for
which finite-size scaling of the entropy and its fluctuations show that the
entanglement phase transition becomes sharp for large systems. Quantum circuits
can describe a very large range of dynamics, while weak measurements provide a
rather general description of environments, where the variable measurement
strength bridges between the fully projective measurements and the fully unitary
dynamics of the unobserved system. We, therefore, anticipate that our findings
generalize to a large range of settings that deserve further investigation. This
includes systems with nonstochastic unitary dynamics as well as different types
of interactions, observables, or dimensions, but also practical applications
aiming at the control of quantum systems via  engineering of their environment.

\begin{acknowledgments}
This research was funded by UK Engineering and Physical Sciences
Research Council (EPSRC) via Grant No.~EP/P010180/1. Computer time was
provided by Lancaster University's High-End Computing facility.
All relevant data present in this publication can be accessed at 
\url{https://dx.doi.org/10.17635/lancaster/researchdata/310}.
\end{acknowledgments}

\appendix

\section{Further analysis of the scaling of entropy and its variance}
\label{app:scaling}

In this work we used the power-law scaling ansatz of Eq.~(\ref{eq:fss})
to fit the entropy dependence on the measurement probability $p$ or measurement
strength $\lambda/\Delta$. This choice of the scaling function can be justified
based on the general constraints for the volume to area law in many-body
localization transition \cite{Grover2014,Khemani2017}.
 In random circuits, a logarithmic scaling
function has been employed to determine the critical values of the measurement
probability for circuits of Clifford algebra operators only~\cite{Li2018,
Li2019, Gullans2019}. As the use of a logarithmic scaling is justified in
Clifford random circuits by the emergence of a conformal symmetry at the
transition~\cite{Li2019}, we here also report the results of the critical value and
exponent from a logarithmic scaling function in our model, even though no conformal
symmetry is known to arise.

\begin{figure}[t]
	\centering
	\includegraphics[width=\linewidth]{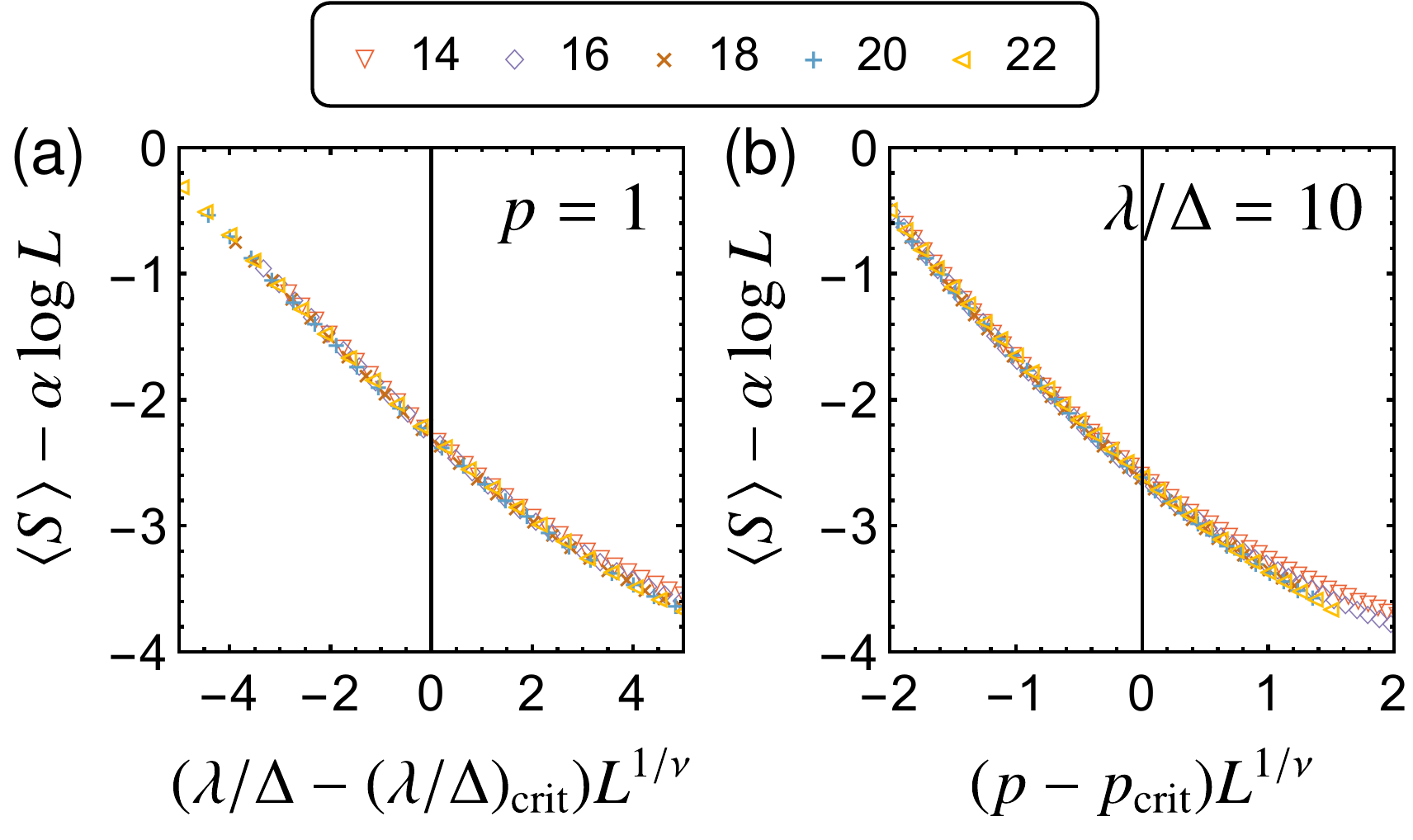}
	
	\caption{Data collapse of the entanglement entropy for (a) $p=1$, and (b)
	$\lambda / \Delta = 10$, assuming a logarithmic scaling.}
	
	\label{fig:log-scaling}
\end{figure}

We used the scaling function
\begin{equation}
  \langle S\rangle - \alpha \ln L = F\left( L^{1/\nu} (x - x_\text{crit})
  \right),
  \label{eq:log_scaling}
\end{equation}
and collapsed the data by varying $\nu$ and $\alpha$ (see
Fig.~\ref{fig:log-scaling}). The corresponding critical values are $(\lambda /
\Delta)_\text{crit} = 0.30(2), \nu = 0.80, \alpha = 1.6$ for the $p = 1$
transition and $p_\text{crit} = 0.14(3), \nu = 0.96, \alpha =
1.7$ for the $\lambda / \Delta = 10$ transition. We note that the critical
values are different than the ones reported in the caption of
Fig.~\ref{fig:p1_entropy}, but the fitting of the scaling function is slightly
worse across the full range of data. A more detailed comparison would require
much larger system sizes, which are not accessible in exact numerics for the
Haar model considered here. Furthermore, given the finite system sizes in our
study, we cannot rule out a crossover from power-law scaling to logarithmic
scaling from the weak-measurement to the strong-measurement transition.

\begin{figure*}[t]
	\centering
	\includegraphics[width=0.45\linewidth]{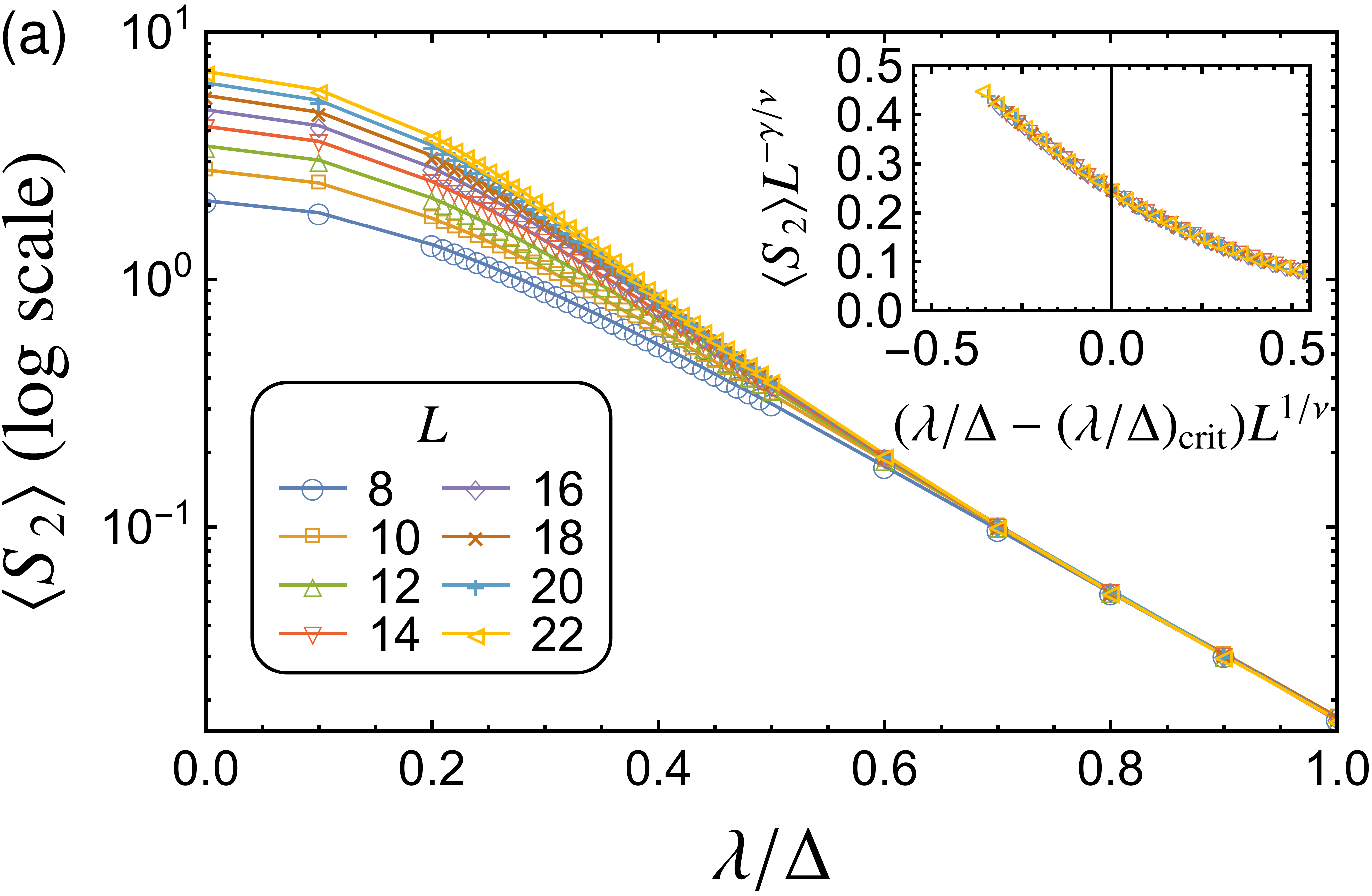}
  \includegraphics[width=0.45\linewidth]{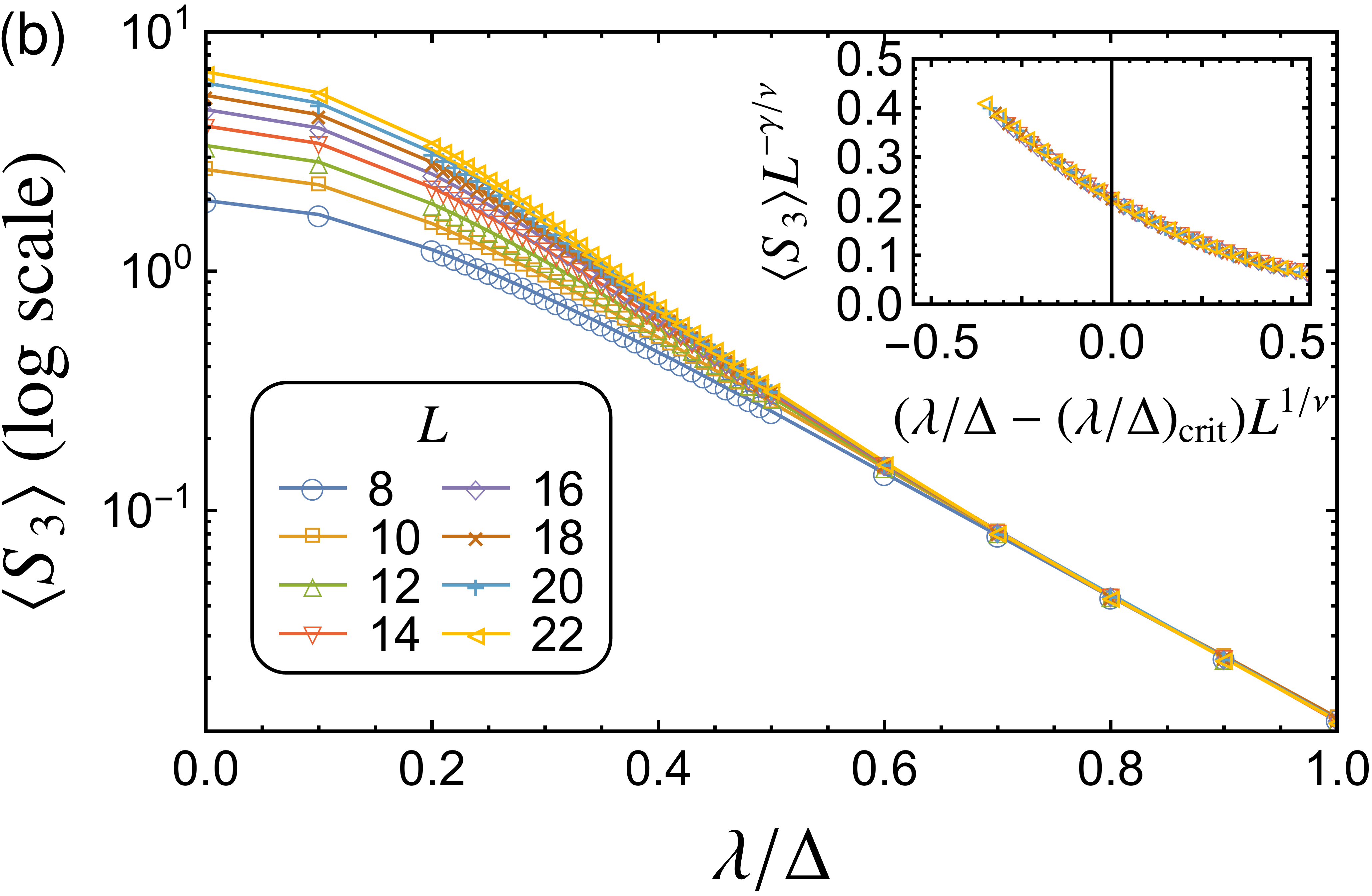}
	\includegraphics[width=0.45\linewidth]{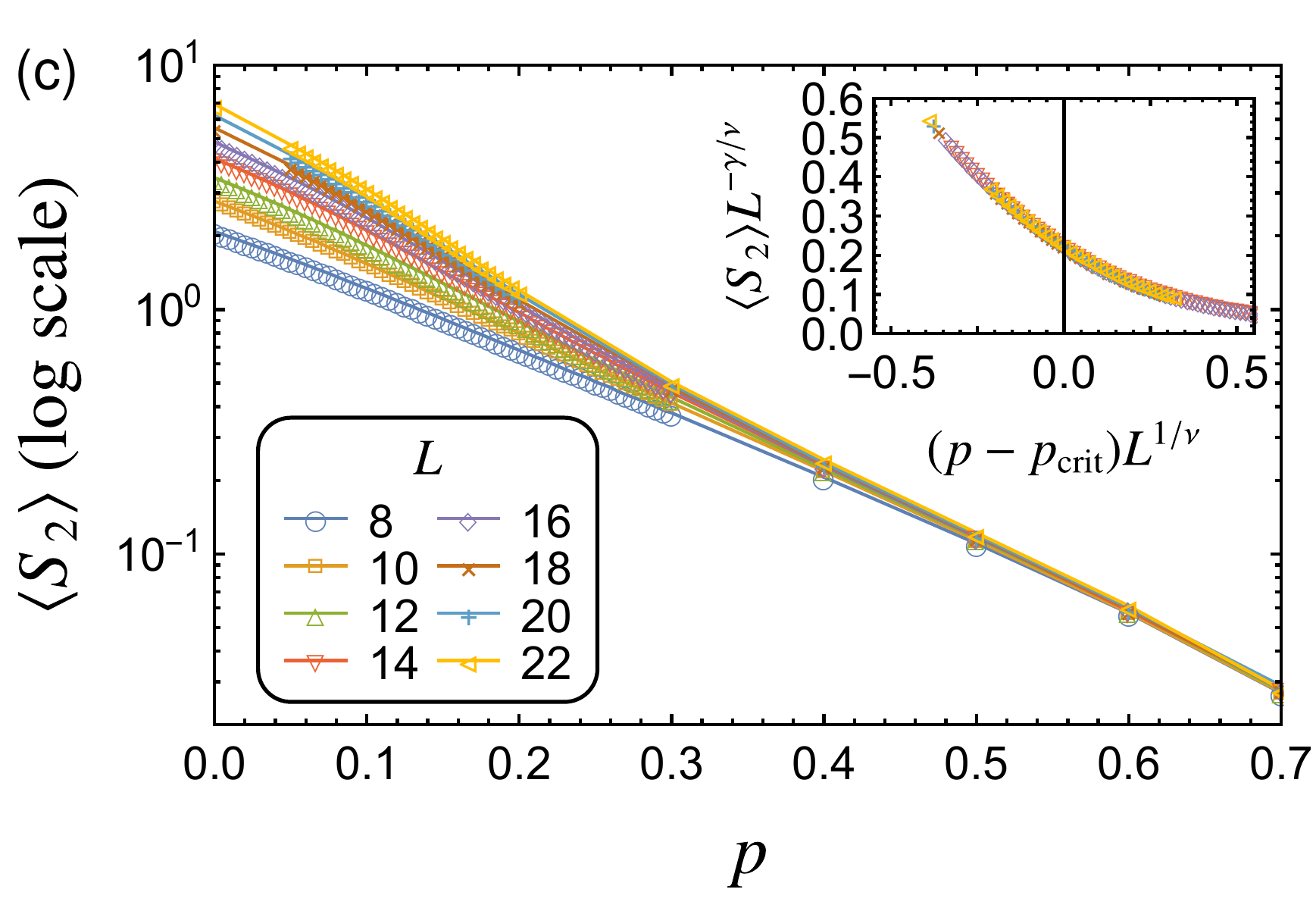}
  \includegraphics[width=0.45\linewidth]{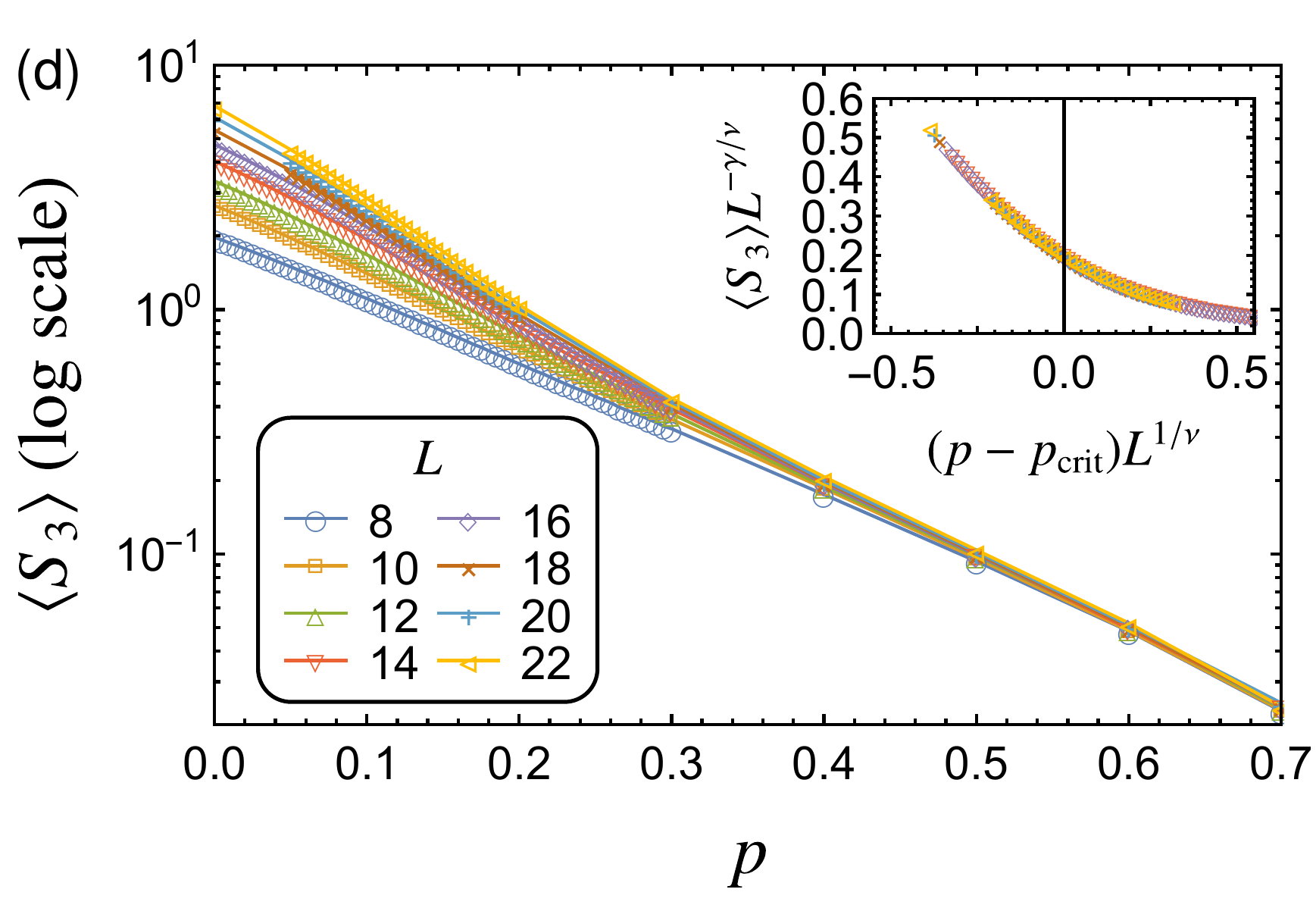}

	\caption{R\'enyi entropies $S_2$ and $S_3$, across the phase transition
	with (a, b) $p = 1$ (where measurements are always performed), and with (c, d)
	$\lambda/\Delta = 10$ (where measurements are always strong). Insets show the
	data collapse assuming that the entropies scale with a power law near the critical
  region. Transition points and the critical exponents are
  (a) $(\lambda/\Delta)_\text{crit} = 0.29(1), \gamma = 1.54, \nu = 2.24$;
  (b) $(\lambda/\Delta)_\text{crit} = 0.29(2), \gamma = 1.55, \nu = 2.27$;
  (c) $p_\text{crit}                = 0.11(1), \gamma = 1.94, \nu = 2.35$;
  (d) $p_\text{crit}                = 0.11(2), \gamma = 2.01, \nu = 2.45$.
  }
	
	\label{fig:renyi}
\end{figure*}

\begin{figure*}[t]
	\centering
	\includegraphics[width=0.45\linewidth]{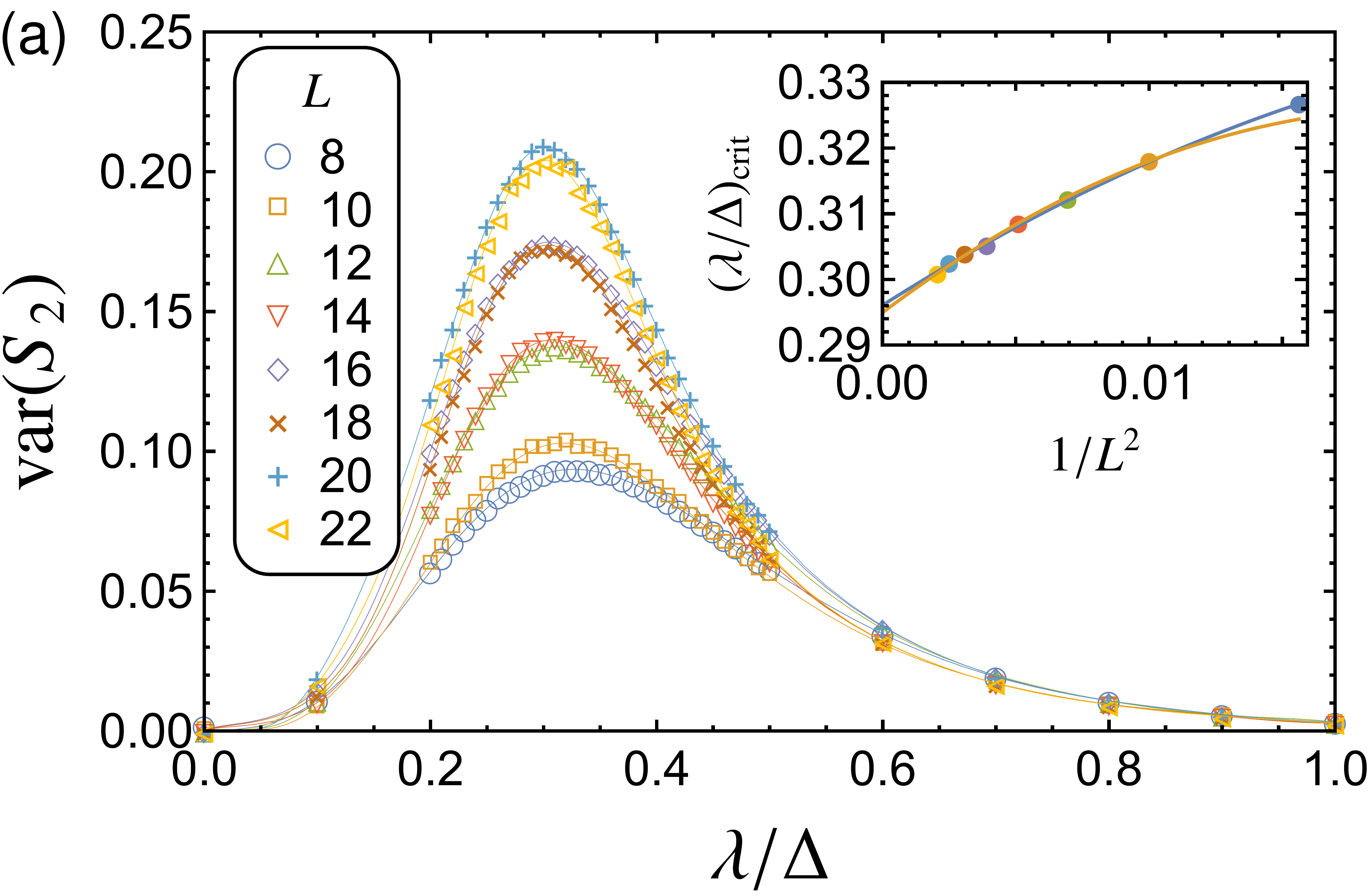}
  \includegraphics[width=0.45\linewidth]{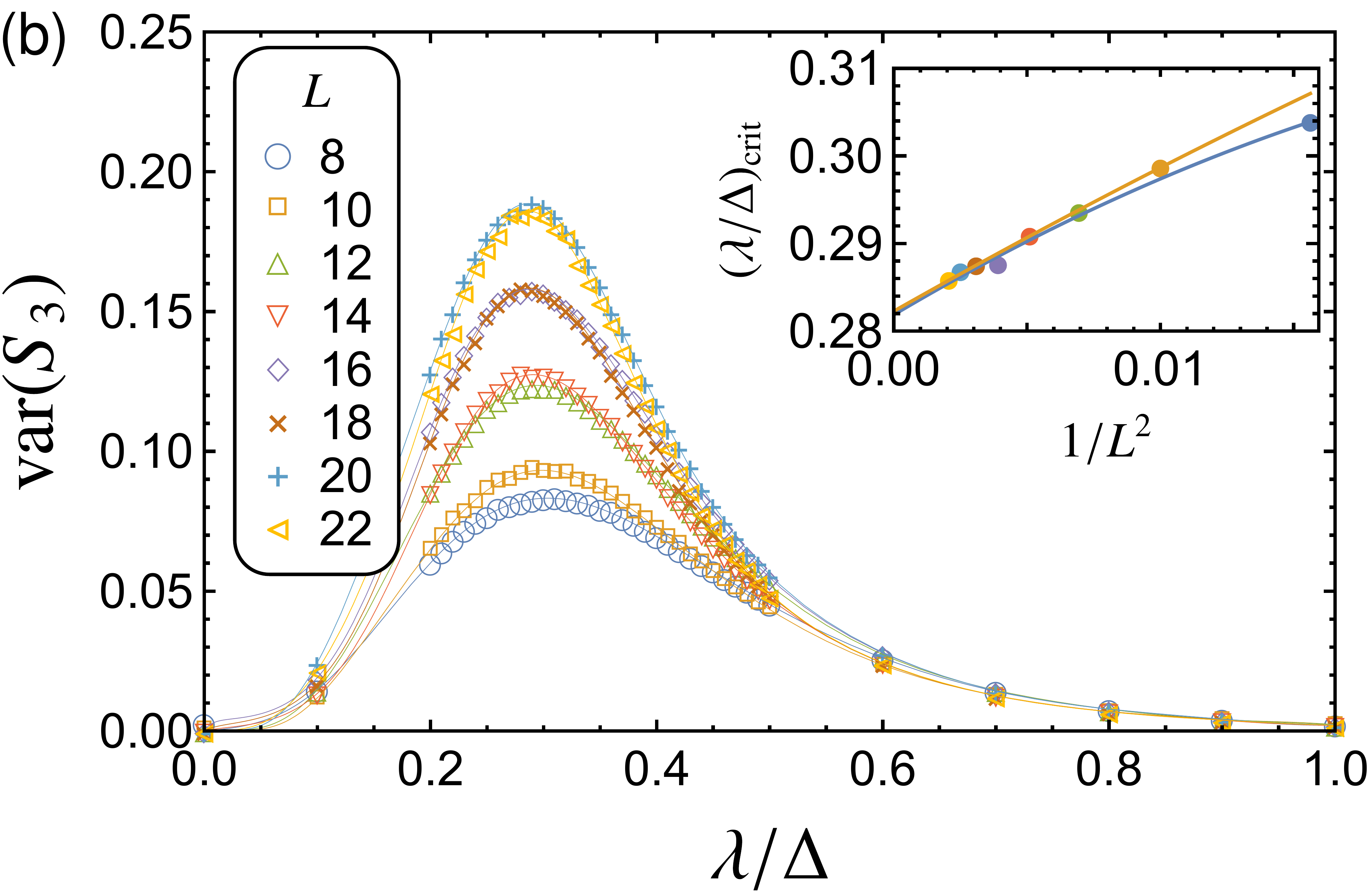}
	\includegraphics[width=0.45\linewidth]{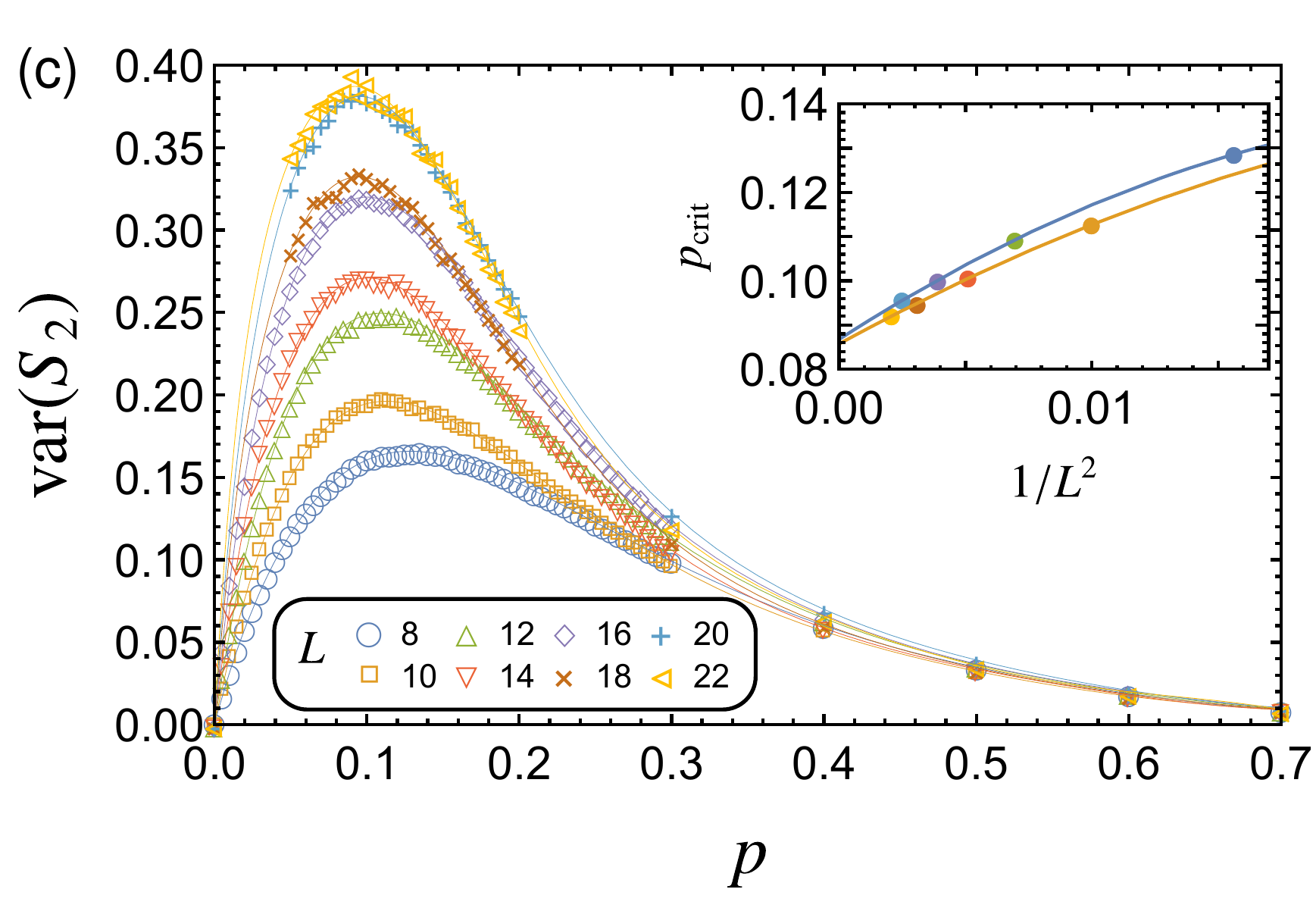}
  \includegraphics[width=0.45\linewidth]{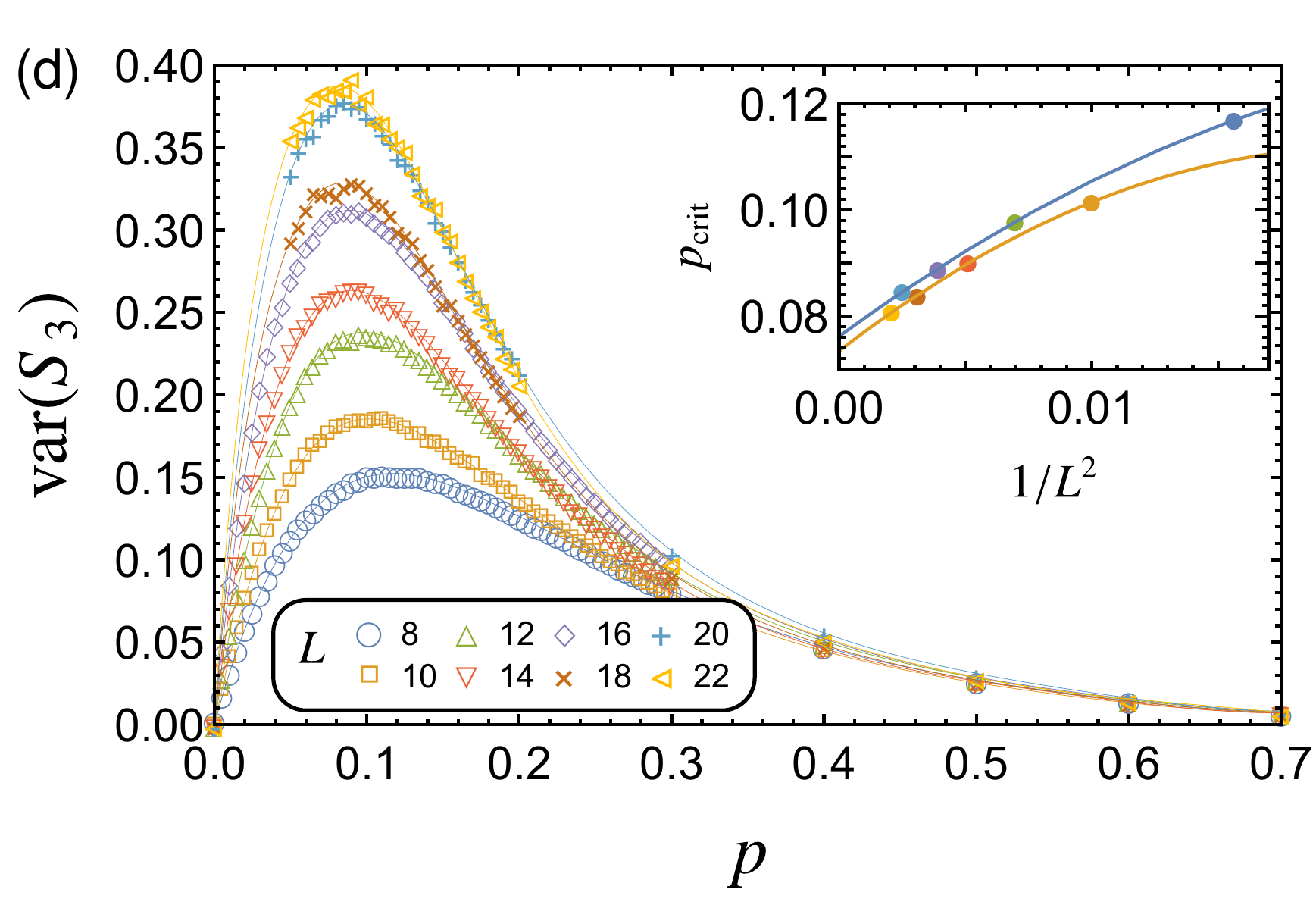}

	\caption{Variance of R\'enyi entropies $S_2$ and $S_3$, across the phase
	transition with (a, b) $p = 1$ (where measurements are always performed), and
	with (c, d) $\lambda/\Delta = 10$ (where measurements are always strong).
	Insets show extrapolation of the variance peak position to infinite system
	size. "Extrapolations give the following estimates of the entanglement 
	transition:
  (a) $(\lambda/\Delta)_\text{crit} = 0.289(2)$,
  (b) $(\lambda/\Delta)_\text{crit} = 0.279(3)$,
  (c) $p_\text{crit}                = 0.0864(5)$,
  (d) $p_\text{crit}                = 0.075(1)$.
  }
	
	\label{fig:renyi_var}
\end{figure*}

In order to corroborate the results obtained from the scaling of the entropy
variance, we have performed similar analysis for the second and third R\'enyi entanglement
entropies, $S_2, S_3$, where
\begin{equation}
  S_\alpha = {1\over 1-\alpha} \ln \text{tr}(\rho_A^\alpha).
\end{equation}
For $\alpha \to 1$ this becomes the von Neumann entropy. The results are reported
in Figs.~\ref{fig:renyi} and \ref{fig:renyi_var}, and show clear signatures of a
phase transition consistent with the picture obtained from the von Neumann
entropy. These entropies are of particular
experimental interest~\cite{Abanin2012, Daley2012, Islam2015}.

\bibliography{refs}

\end{document}